\documentclass[]{aa}
\usepackage{txfonts}
\usepackage{epsfig}
\usepackage{graphicx}
\usepackage{natbib}
\usepackage{url}
\usepackage{twoopt}
\usepackage[breaklinks=true]{hyperref}
\usepackage{dcolumn}

\usepackage{color}
\usepackage{ulem}

\begin{document}

\title{Line shapes of the magnesium resonance  lines  in cool DZ 
 white dwarf atmospheres}
\author{N.~F.~Allard    \inst{1,2} 
   \and J.~F.~Kielkopf  \inst{3} 
  \and S. Blouin \inst{4} 
   \and P. Dufour \inst{4}
   \and F.~X.~Gad\'ea   \inst{5}
   \and T.~Leininger    \inst{6}
   \and   G. Guillon    \inst{7}
 }

\offprints{nicole.allard@obspm.fr}

\institute{
GEPI, Observatoire de Paris,  Universit\'e PSL, UMR 8111, CNRS, 
61, Avenue de l'Observatoire, F-75014 Paris, France \\
\email{nicole.allard@obspm.fr}
\and
Sorbonne Universit\'e, CNRS, UMR7095, Institut d'Astrophysique de Paris, 
 98bis Boulevard Arago, PARIS, France \\
\and
        Department of Physics and Astronomy, 
           University of Louisville, Louisville, Kentucky 40292 USA  \\
 \and
D\'epartement de physique, Universit\'e de Montr\'eal, Montr\'eal,
Qu\'ebec H3C 3J7, Canada \\
\and
        Department of Physics and Astronomy, 
           University of Louisville, Louisville, Kentucky 40292 USA  \\
\and
Laboratoire de Physique et Chimie Quantique, UMR 5626, Universit\'e de      
           Toulouse (UPS) and CNRS, 118 route de Narbonne, 
           F-31400 Toulouse,   France \\
\and
             Laboratoire Interdisciplinaire Carnot de Bourgogne, 
         UMR6303, CNRS, Universit\'e de Bourgogne Franche Comt\'e, 
         21078 Dijon Cedex, France \\
}
\date{received 10 august 2018; accepted 26 august 2018}

\abstract
{Line shapes of the magnesium resonance lines in  white dwarf 
spectra are determined by  the properties of magnesium atoms
 and the structure of the white dwarf atmosphere. 
Through their blanketing effect, these
lines have a dominant influence on the model structure and thus on the 
determination from the spectra of other physical parameters that describe the 
stellar atmosphere and elemental abundances. 
}
{In continuation of previous work on Mg$^+$He lines in the UV, we present
theoretical profiles of the resonance line of neutral Mg perturbed by He at the
extreme density conditions found in the cool largely transparent atmosphere 
of DZ white dwarfs.}
{We accurately determined the broadening of Mg by He in a unified theory of collisional line profiles
using   ab initio calculations of MgHe
potential energies and transition matrix elements among the singlet
 electronic states that are involved for the observable spectral lines. }
{We  computed the shapes and line parameters of the  Mg lines
 and studied their dependence on helium densities and temperatures.
We present results over the full range of  temperatures from 4000 to 12000~K
needed for input to stellar spectra models.
Atmosphere models were
constructed for a range of effective temperatures and surface gravities
typical for cool DZ white dwarfs. We present synthetic
spectra tracing the behavior of the Mg resonance line profiles under the
low temperatures and high gas pressures prevalent in these atmospheres.}
{The determination of accurate opacity data of magnesium resonance lines
  together with an improved  atmosphere model code lead to a good
  fit of cool DZ white dwarf stars. The broadening of spectral
  lines by helium needs to be understood to accurately determine the
 H/He and Mg/He abundance ratio in DZ white dwarf atmospheres.
  We emphasize that no free potential parameters or  ad hoc adjustments
    were used to calculate the line profiles.
}


\authorrunning{N.~F.~Allard et al.}
\titlerunning{Mg lines in the UV}

\maketitle

\section{Introduction}
\begin{figure}
\resizebox{0.46\textwidth}{!}
{\includegraphics*{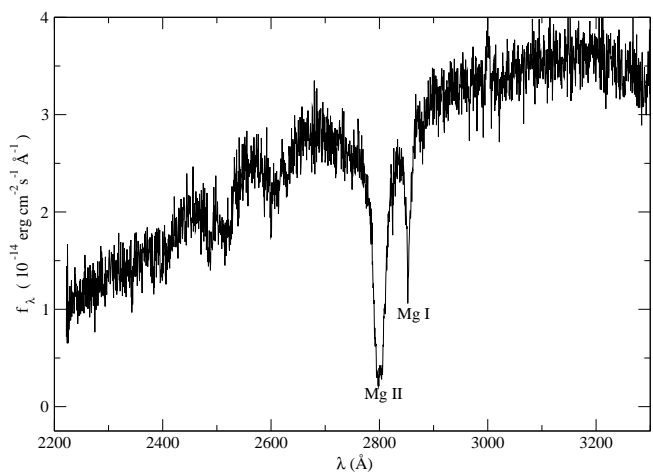}}
\caption  {Ultraviolet spectrum of the DZ white dwarf L745-46A. The distinctive resonance lines of Mg at 2852~\AA\/ and Mg$^+$
  at 2796/2803~\AA\ are clearly visible\/
\citep{hst2593S}. \label{fig:l745}}
\end{figure}

The UV spectra of cool DZ white dwarfs, which are rich in helium, show the 
resonance lines of Mg at 2852~\AA\/ and Mg$^+$ at 2796/2803~\AA\/.
 An example in Fig.~\ref{fig:l745} shows the ion and neutral 
lines in a spectrum of L745-46A.
The resonance-broadened wings extend and decrease monotonically very far  
on the red long-wavelength  side of line  center.  
Their blue wings, however,  show ``satellite bands'', features 
 that are due to the absorption  of radiation during the Mg-He and 
Mg$^+$-He collisions, which are close to 2750~\AA\/ and 2300~\AA \ for the neutral atom and the ion, respectively.
The aim of this study is to clarify the contribution of each of these two 
lines in their far wings to stellar opacity.

\begin{figure}
   \resizebox{0.46\textwidth}{!}{\includegraphics*{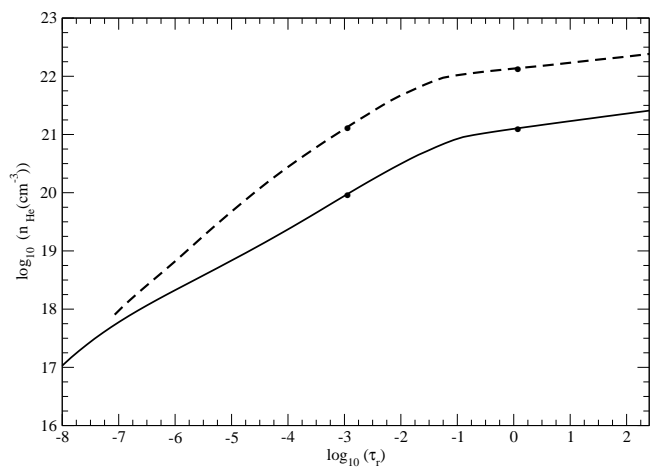}}
\caption  { He density as a function of Rosseland mean
  optical depth assuming  atmospheric parameters used in synthetic spectra 
  studied  in Sect.~\ref{sec:blanket}.
  $T_{\rm eff}$ = 8000~K (full line), $\log\,{\rm Ca/He} = -9$,
                                    $\log\,{\rm Mg/He} = -7.3$,
                                    $\log\,{\rm H/He} = -3.5$.
  $T_{\rm eff}$=6000~K (dashed line), $\log\,{\rm Ca/He} = -10.5$,
                                     $\log\,{\rm Mg/He} = -7.3$, and
                                     $\log\,{\rm H/He} = -3.5$.
\label{TandRhovsTauR}}
\end{figure}

\citet{blouin2018} have now developed an improved atmosphere model code to 
accurately describe cool DZ white dwarfs taking into account
non-ideal high-density effects arising at the photosphere.
The line profiles of the
resonance lines of Mg$^+$  \citep{allard2016c} and neutral Mg described
in this paper have been  included.
Figure~\ref{TandRhovsTauR} shows a typical pressure structure for a
helium-rich white dwarf at $T_{\rm eff} = 6000K $, in which the helium
density can reach $10^{22}$ cm$^{-3}$.
 In these physical conditions, the emitting atoms can experience multiple 
simultaneous energetic perturbations from the surrounding helium atoms. 
Recently, \citet{reggami2011} calculated the collisional  broadening of
 Mg for low helium density  in the  wavelength range from
 2600 to 3100~\AA\/ and for a temperature range from  100 to 3000~K.
 The higher temperature of white dwarfs makes the close
 collisions more likely, which
 are affected by short-range atomic interactions, while  the higher density
 increases the probability of all collisions and their effects on
 the spectral line shape compared to, for example, Doppler broadening.

Since our review paper 36 years ago  \citep{allard1982}, 
considerable progress in unified line broadening theory and in computational 
technology
 now enables us to calculate neutral atom spectra given the potential 
energies and
radiative transition moments for relevant states
of the radiating atom interacting with other atoms in its environment.  
Although our unified theory has been developed in \citet{allard1999},
 and a detailed discussion is presented there, we provide an overview
Sect.~\ref{sec:theory} for its use in this context. 
Because the underpinning
atomic physics is understood, theoretical models may be drastically
improved compared to previous work such as 
\citet{koester2000} and \cite{wolff2002} by including a more complete
 representation of the interactions,
especially in the region of atomic separations that determine the line 
wings. 
Nevertheless, the  main limitation remains a lack of precision 
in theoretical fundamental
atomic interaction data suitable for such spectroscopic calculations. 
 In a field
where most work focuses on lower precision for use in chemistry and 
reaction kinetics,  it is fortunate that  Mg-He and  Mg$^+$-He molecular data 
are now very well studied and have accurate asymptotic energies
available. The potentials we use are described in detail
in Sect.~\ref{sec:potentials}.
In  Sect.~\ref{sec:wing} we examine  the far wings
quantitatively for the MgHe and Mg$^+$He lines, which both contribute to the ultraviolet spectra. 
We  present  line profiles obtained over the full range of temperatures from
4000 to 12000~K for helium densities varying from
$10^{21}$ to $10^{22}$ cm$^{-3}$. 
In Sect.~\ref{sec:blanket} we study the relative contribution of the 
two  resonance lines 
in their far wings and how they can contribute to the line blanketing.
At sufficiently low densities of  perturbers, the symmetric center 
of a spectral line is Lorentzian and can be defined by two line 
parameters, the width and the shift of the main line.
 The impact approximation determines the asymptotic 
  behavior of the  unified line shape autocorrelation function. 
The Lorentzian width  can be readily extracted, and is presented in 
Sect.~\ref{sec:parameter}.
Finally, we report the study of WD2216-657 in Sect.~\ref{sec:astro}. 

\section{General expression for the spectrum in an adiabatic representation
\label{sec:theory}}

In a unified treatment, the complete spectral 
energy distribution is computed from the core to the far line wing.
The semiclassical methods used by \citet{sando1973} were generalized to apply
to the entire profile in the unified Franck-Condon theory of \citet{szudy1975}. 
This method has been widely used for the analysis of experiments on line
 broadening \citep{allard1982}. Its simplicity derives from a  neglect, in 
the line wing, of multiple close collisions and non-adiabatic effects, and 
from the use of a cubic expansion of the phase shift of the WKB wavefunctions.
The profiles computed with this approach provide a unified expression from
 the line core to the wing. However, in the satellite region of the line 
wing, the results are only valid when densities are so low that the 
probability of close collisions of more than two atoms is negligibly small.

With $n_{\mathrm{He}}$ densities  above 
$10^{21}$ cm$^{-3}$ for Ross 640 \citep{blouin2018},
reaching 2$\times$ $10^{22}$ cm$^{-3}$ for vMa2
(Dufour, private communication),
 multiple perturber effects have to be taken into account.
Each one of these white dwarfs constitutes a  very interesting case study.
In dense plasmas, as in these very cool DZ white dwarfs, a
 reliable determination of the line profiles 
 that is applicable in {\it all} parts of the line at  {\it all}
  densities
  is the Anderson semi-classical theory~\citep{anderson1952}, which uses
  the Fourier transform  of an autocorrelation function.
A unified theory of spectral line broadening \citep{allard1999}
has been developed to calculate neutral atom spectra given the
interaction and the radiative transition moments of relevant states
of the radiating atom with other atoms in its environment.
Our approach is based on the quantum theory of
spectral line shapes of~\citet{baranger1958a,baranger1958b} 
 developed in
an adiabatic representation to include the degeneracy of atomic
levels~\cite{royer1974,royer1980,allard1994}. 
 The spectrum $I(\Delta\omega)$ can be written as the 
Fourier transform (FT) of the dipole autocorrelation function $\Phi(s)$ ,
\begin{equation}
I(\Delta\omega)=
\frac{1}{\pi} \, Re \, \int^{+\infty}_0\Phi(s)e^{-i \Delta\omega s} ds,
\label{eq:int}
\end{equation}
where $s$ is time.
The FT in Eq.~(\ref{eq:int}) 
is taken such that $I(\Delta$$\omega)$ is
normalized to unity when integrated over all frequencies,
and $\Delta$$\omega$ is measured relative to  the 
unperturbed line.
A pairwise additive assumption
 allows us to calculate the total profile $I(\Delta \omega)$ where
all the perturbers interact as the FT of the $N^{\rm th}$ power of the
autocorrelation function $\phi (s)$ of a unique atom-perturber
pair. Therefore
\begin{equation}
\Phi(s)=(\phi(s))^{N}\; .
\end{equation}
That is to say, we neglect the interperturber correlations. The radiator
can interact with several perturbers simultaneously, but the perturbers do
not interact with each other. This is what~\citet{royer1980} 
calls the totally uncorrelated perturbers approximation.
The fundamental result expressing the autocorrelation
function for many perturbers in terms
of a single perturber quantity $g(s)$ was first obtained
 by~\citet{anderson1952} and~\citet{baranger1958a} 
in the classical and quantum cases, respectively.
From the point of view of a general classical theory, the solution to
the ~\citet{anderson1952} model corresponds 
to the first-order approximation
in the gas density obtained by the cumulant expansion method~\citep{royer1972}.
The higher-order terms representing correlations 
between the perturbers  are neglected since they  are
 extremely complicated~\citep{royer1972,kubo1962,kubo1963,vankampen1974}.
We  obtain for a perturber density n$_p$ 
\begin{equation}
\Phi(s) = e^{-n_{p}g(s)}.\end{equation}
The decay of the autocorrelation function $\Phi (s)$ with time $s$ leads to
atomic line broadening. It depends on the density of perturbing atoms
$n_p$ and on their interaction with the radiating atom.
The molecular potentials and radiative dipole transition moments are input data
for a unified spectral line shape evaluation.
The dipole autocorrelation function $\Phi (s)$ is evaluated for
a classical collision path with an average over all possible collisions.
For a transition \mbox { $\alpha =(i,f)$} from an initial state~$i$ 
to a final state~$f$, we have 
\begin{eqnarray}
g_{\alpha}(s) && \,= \frac{1}
{\sum_{e,e'} \, \! ^{(\alpha)} \, |d_{ee'}|^2 }
\sum_{e,e'} \, \! ^{(\alpha)} \; \; \nonumber \\ 
&&  \int^{+\infty}_{0}\!\!2\pi b db
\int^{+\infty}_{-\infty}\!\! dx \; 
\tilde{d}_{ee'}[ \, R(0) \, ] \, \nonumber \\ 
&&[e^{\frac{i}{\hbar}\int^s_0 \!\! dt   \,
V_{e'e }[R(t)] } \,
\, \tilde{d^{*}}_{ee'}[R(s)] - \tilde{d}_{ee'}
[R(0)] \, ]. 
\label{eq:gcl}
\end{eqnarray}
In Eq.~(\ref{eq:gcl}), $e$ and  $e'$  label
the energy surfaces  on which the interacting
atoms  approach the initial and final atomic states of the transition 
 as \mbox{ $R \rightarrow \infty$ }.
 The sum $\sum_{e,e'} ^{(\alpha)}$ is
 over all pairs ($e,e'$)  such that
\mbox{$\omega_{e',e}(R) \rightarrow \omega_{\alpha}$} as 
\mbox{$R \rightarrow \infty$}.
We define 
$\tilde{d}_{ee'}(R(t))$ as a  modulated dipole 
(Allard et al.~1999)
\begin{equation}
D(R) \equiv \tilde{d}_{ee'}[R(t)] = 
d_{ee'}[R(t)]e^{-\frac{\beta}{2}V_{e}[R(t)] } \; , \;
\label{eq:dip}
\end{equation}
where $\beta $ is the inverse temperature ($1/kT$).
Here $V_e$ is
the  ground-state potential when we consider absorption profiles, 
or an excited state for the calculation of a profile in emission.
Over regions where $ V_e (r) <0 $, 
the factor $e^{-\beta V_e (r)}$ accounts for bound states of the
radiator-perturber pair, but in a classical approximation wherein the
discrete bound states are replaced by a continuum; thus any band
structure is smeared out. We have for the phase term in  Eq.~(\ref{eq:gcl})

\begin{equation}
\eta(s)=  \frac{i}{\hbar}\int^s_0 \, dt \;
V_{e'e }[ \, R(t) \, ]
\label{eq:phase}
,\end{equation}
 where $\Delta V(R)$, the difference potential, is given by 
\begin{equation}
\Delta V(R) \equiv V_{e' e}[R(t)] = V_{e' }[R(t)] - V_{ e}[R(t)] \; ,
\label{eq:deltaV}
\end{equation}
 and represents the difference between the electronic energies
of the quasimolecular transition. The potential energy for a state $e$ is 
\begin{equation}
V_{e}[R(t)] = E_e[ R(t) ]-E_e^{\infty} \; .
\label{eq:V}
\end{equation}

  At time $t$ from the point of closest approach 
\begin{equation}
R(t) = \left[\rho ^2 + (x+\bar{v} t)^2 \right]^{1/2} \; , \;
\end{equation}
with $\rho$ the impact parameter of the perturber trajectory and 
$x$ the position of the perturber along its trajectory at time
$t=0$.

\begin{figure}
\resizebox{0.46\textwidth}{0.4\textheight}
{\includegraphics*{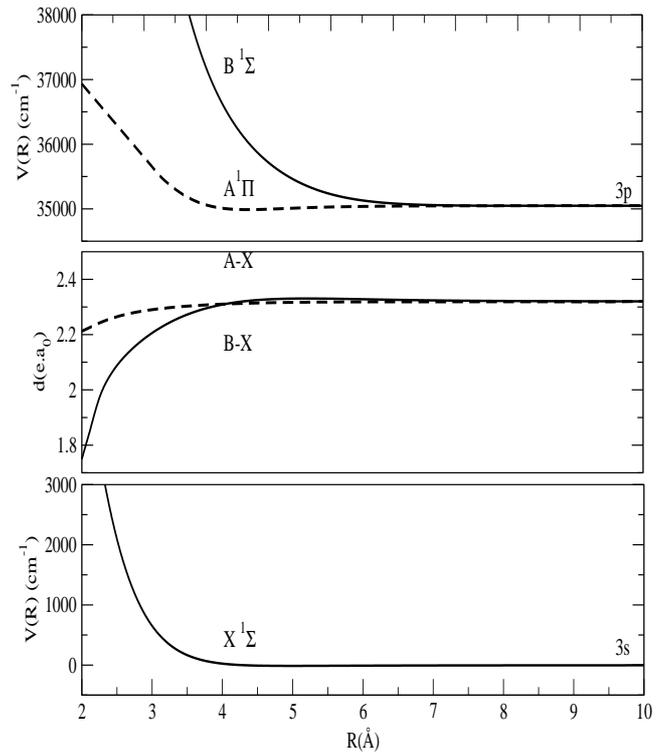}}
\caption  {Potential energies for the  $X$, $A,$ and  $B$ 
 states of the Mg--He molecule.  
$3s$ $X$ $^1\Sigma$ (full line), 
$3p$ $B$ $^1\Sigma$ (full line), 
$3p$ $A$ $^1\Pi  $ (dashed  line), and transition dipole moments 
$B$-$X$ (full line),  $A$-$X$ (dashed  line).\label{potMgHe}}
\end{figure}

\section{MgHe  diatomic potentials \label{sec:potentials}}

 \citet{paulkwiek1998} obtained adiabatic
potential curves of MgHe obtained by means of  pseudo-potentials, 
 $l$ -dependent core polarisation potential (CPP) and single- and double-configuration interaction (SDCI) techniques.
 The semi-empirical Hartree Fock Damped dispersion ansatz (HFD)
 methods have been used in \citet{reho2000} and   accurate
 ab initio calculations for the Mg–He and Ca–He van der Waals
 potential energy curves of the ground state  
have been presented in \citet{hinde2003}. \citet{mella2005} presented 
ab initio calculations for  Mg–He focused on solvation effects of Mg in 
He clusters.

 The most recent calculations  of Mg--He
singlet  potential energies have been 
achieved by Alekseev (2014)~\footnote
{https://www.researchgate.net/profile/Vadim-Alekseev/publications},
 but in a limited  range, $R \sim$ 2.5 to 20 a.u., where $R$ 
denotes the internuclear distance between
the radiator and the perturber. The ab initio potentials computed with 
MOLCAS \citep{aquilante2010}
are described in \citet{hollands2017}.

The potentials we used were computed by ab initio approaches with the MOLPRO 
package \citep{molpro2012}. As for the triplet states \citep{allard2016a}, a pseudo-potential with large core was used for 
Mg  \citep{fuentealba1983},
 and core valence correlation effects were taken into account with the operatorial
 CPP. A large basis set (10s, 9p, 6d, 3f,3g) 
was optimized for Mg, starting from one involving quite high Rydberg states 
\citep{khemiri2013}, together with the rather large  He one of 
\citet{deguilhem2009}. 
MRCI  calculations, with single and double excitations, 
were performed after a CASSCF with four electrons in six orbitals (CAS 4,6)
 to obtain both the energies and the transitions dipole moments. 
A dense grid of internuclear distances has been considered, ranging between 
4 and 500 a.u., in order to cover a broad range from dissociation 
to the shortest distances that are energetically accessible.

The line profile depends only on the two  individual  
transitions  
\mbox {$ \, 3s \, X ^1\Sigma \rightarrow   \, 3p \, B^1\Sigma$}
and 
\mbox {$ \, 3s \, X ^1\Sigma \rightarrow   \, 3p \, A^1\Pi$}. 
Figure~\ref{potMgHe} shows the potentials for the $X$, $A,$ and $B$ states 
and the transition dipole moment for the 
$B$-$X$ and   $A$-$X$ transitions.
For Mg$^+$--He, we used the recent ab initio calculations presented in 
\citet{allard2016c}.

In the next section, we evaluate  collisional profiles 
 for relevant temperatures and
densities that are appropriate for modeling  He-rich white dwarf stars. 

\begin{figure}
\resizebox{0.46\textwidth}{!}{\includegraphics*{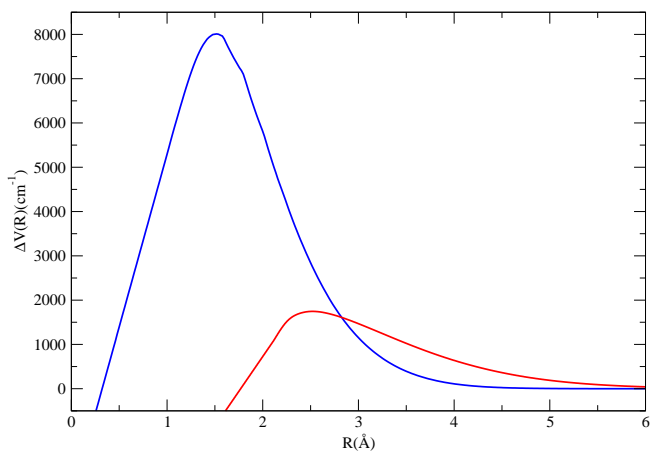}}
\caption{ $\Delta V$ for the  $B$-$X$ transition of the MgHe line (red curve), 
  compared to the Mg$^+$He  line (blue curve), see Fig.~4 of
  \citet{allard2016c}.
\label{potdipBX}}
\end{figure}

\section{Temperature and density dependence of the Mg lines
\label{sec:temperature}}

Figure~\ref{TandRhovsTauR} shows a typical pressure
structure corresponding to the synthetic spectra studied 
in Sect.~\ref{sec:blanket}. This was obtained
  with the atmosphere model code described in \citet{blouin2018}.

\subsection{Study of the line wings \label{sec:wing}}

\begin{figure}
\resizebox{0.46\textwidth}{!}{\includegraphics*{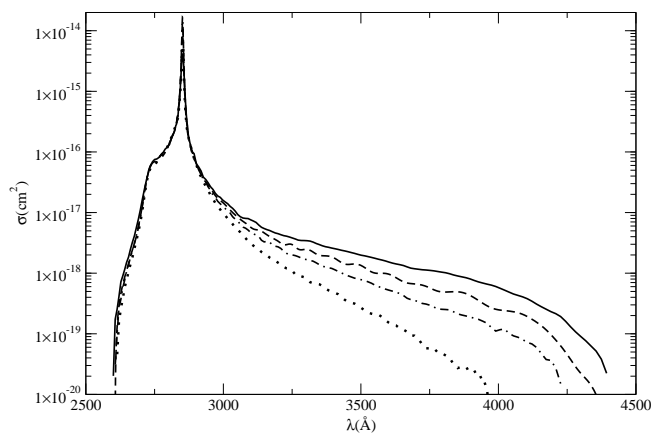}}
  \caption{Variation of the MgHe line profile with temperature
(from top to bottom,  \mbox{$T$  =12000, 8000, 6000, and 4000~K})
  for $n_{\mathrm{He}} = 10^{21}$ cm$^{-3}$
 \label{satvarTMgHe}}
\end{figure}

\begin{figure}
\resizebox{0.46\textwidth}{!}{\includegraphics*{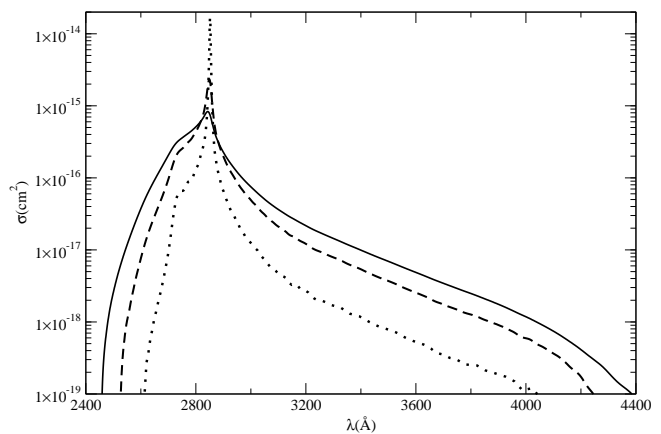}}
\caption  {Theoretical absorption cross-sections of the MgHe line. 
The  density of the perturbers  is $n_{\mathrm{He}} = 10^{22}$, 5$\times$ $10^{21}$
 and $10^{21}$ cm$^{-3}$, from top to bottom.
The temperature is 6000~K.
\label{profdens21}}
\end{figure}

 The prediction of the shape  of the \mbox{$3s \rightarrow 3p$} line
 requires studying  the potential energy difference 
$\Delta V(R)$ in Eq.~(\ref{eq:deltaV}).
 The unified theory predicts that line  satellites  will be centered periodically at 
frequencies corresponding to integer multiples of
the extrema of  $\Delta V$. 
Figure~\ref{potdipBX} shows $ \Delta V(R)$ for the 
 \mbox {$ \, 3s \, X ^1\Sigma \rightarrow   \, 3p \, B^1\Sigma$} transition,
 which leads
 to the formation of satellites on the Mg resonance line. 
The maximum in $\Delta V$ in this case occurs at larger
 internuclear distances ($R$ $\sim$ 2.5~\AA\/) than for the Mg$^+$--He molecule.
As a consequence, the average number of perturbers in the interaction 
volume is larger, leading to a higher probability of multiple pertuber
 effects, and making a second satellite strong enough to appear distinctly
 as a shoulder in Fig.~\ref{satvarTMgHe}.
 When the density 
reaches $5 \times 10^{21}$~cm$^{-3}$, the shoulder completely blends 
with the first line satellite seen in Fig.~\ref{profdens21}. 
 We have known for a long time, since the pioneering work
of~\citet{mccartan1969}, that multiple satellites can be observed
experimentally at very high densities. 
A definitive observation of
multiple perturber satellites was reported in~\citet{kielkopf1979}, and 
the effects have been seen in many cases since then.

Collisional line profiles are evaluated for 
\mbox{$n_{\mathrm{He}}$=1$\times$~10$^{21}$ cm$^{-3}$} 
from 4000 to 12000~K.
Furthermore, the  change of  radiative transition moment $D(R)$ 
(Eq~(\ref{eq:dip})
 with atomic separation
has a significant  effect on the line wing extension according to the 
effective temperatures. 
These effects have heretofore not been studied  in UV spectra of cool 
white dwarfs using  accurate molecular data. 
 The blue line
 wings shown in (Fig.~\ref{satvarTMgHe}) are almost unchanged with increasing 
temperature, whereas the red wings extended very far.

 \citet{reggami2011} analyzed the line satellite structure 
quantum-mechanically. 
They constructed two different sets 
of potentials. In their  Set~1 they used data of \citet{hinde2003}
 for the ground state and \citet{mella2005} and \citet{reho2000} for the 
excited states.
In their Set~2 they used \citet{paulkwiek1998} data.
 On the blue side they obtained satellite structure for T $>$ 1800~K, 
 located around 2760~\AA\/ when Set~1 is used. 
This satellite position is close to 
ours.  For the other set of potentials the position of the satellite is
around 2720~\AA\/, and the comparison in
Figs.~\ref{satvarTMgHe}-\ref{profdens21} stresses the
 importance of
accurately determining the atomic interaction potentials
when the far wings of these lines are to be evaluated.

\subsection{Blanketing effects of the Mg lines\label{sec:blanket}}

\begin{figure}
\resizebox{0.46\textwidth}{!}
{\includegraphics*{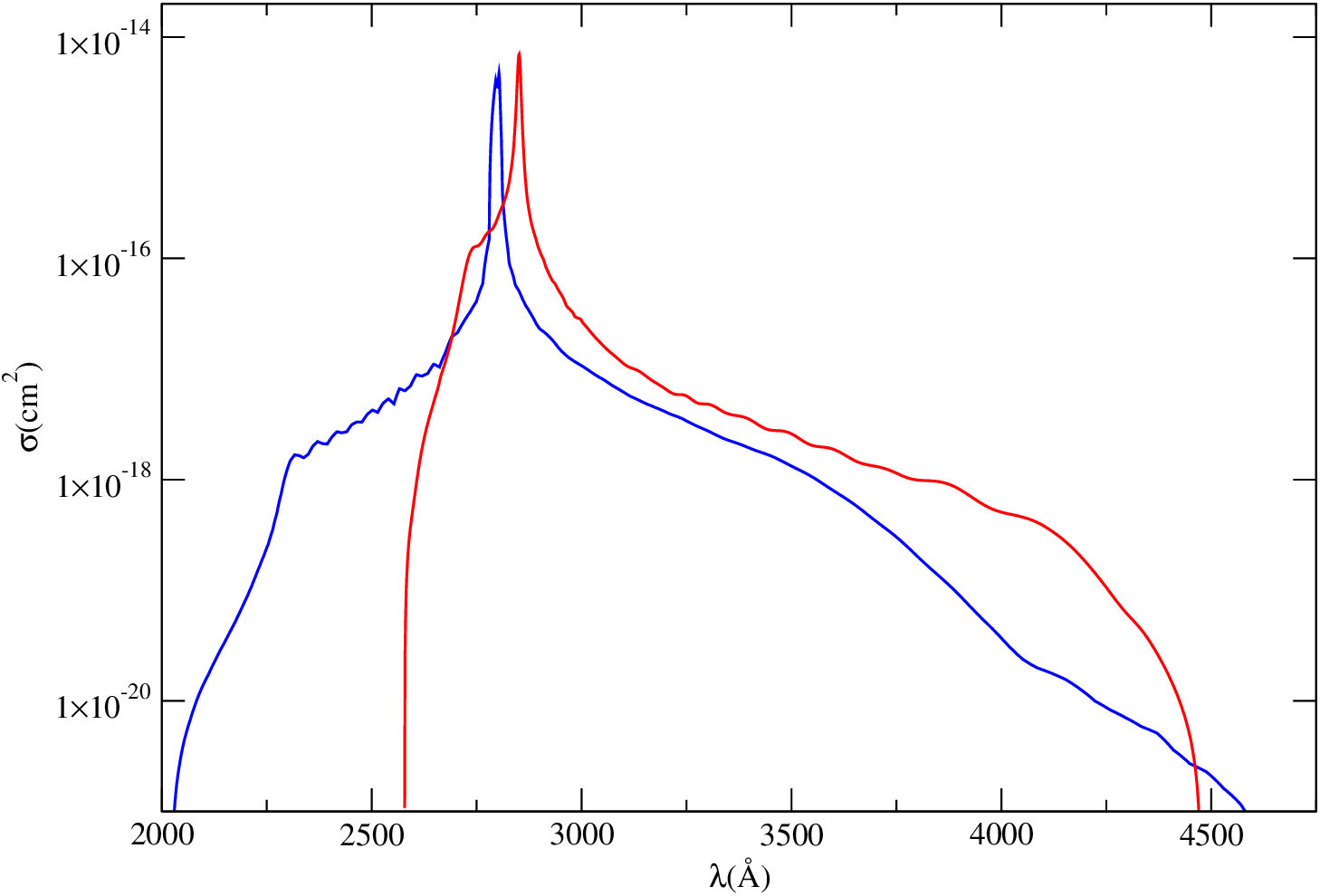}}
  \caption{Comparison  of the MgHe line (red line) to the sum of the 
two components of the Mg$^+$He  line (blue line) 
(\mbox{$T$  =8000~K} and  $n_{\mathrm{He}}$ = 2$\times$~$10^{21}$ cm$^{-3}$).
 \label{sig8000MgHeMgIIHe}}
\end{figure}

For the Mg--He pair, the maximum $\Delta V_{\rm{}max}$ is 
 smaller, 1756 cm$^{-1}$
versus 8100 cm$^{-1}$ for the Mg$^+$-He pair (Fig.~\ref{potdipBX}). 
The MgHe line satellite is then closer  to the main line than 
the Mg$^+$He satellite, as shown in 
Figs.~\ref{sig8000MgHeMgIIHe} and ~\ref{sig6000MgHeMgIIHe}.
For a helium  density  $> 5\times$~$10^{21}$ cm$^{-3}$ , the  red wings
have a shoulder-like structure, and the
satellite structures become less prominent. The blue wings fall off very
rapidly.

\begin{figure}  
\centering  
\resizebox{0.46\textwidth}{!}{\includegraphics*{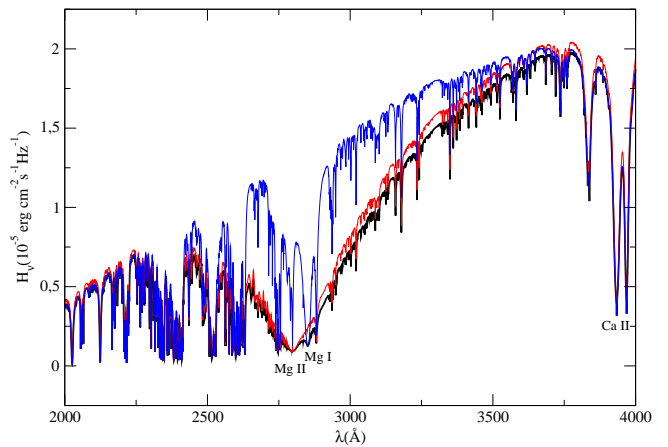}}
\caption  {Synthetic spectrum computed at
  $T_{\rm eff} = 8000~K$, $\log\,{\rm Ca/He} = -9$, $\log\,{\rm Fe/He} = -8.4$,
  $\log\,{\rm Mg/He} = -7.3$, and $\log\,{\rm H/He} = -3.5$, compared to
  synthetic spectra
  computed without the MgII (in blue) and the MgI (in red) line opacity.
\label{MgUV8000}}
\end{figure}

Using the  stellar atmosphere code of \citet{blouin2018}, we constructed
atmosphere models 
 for a range of effective temperatures and surface gravities
typical for cool DZ white dwarfs. We present synthetic
spectra tracing the behavior of the Mg resonance line profiles under the
low temperatures and high gas pressures prevalent in these atmospheres.
Figures~\ref{MgUV8000} and ~\ref{MgUV6000} show the relative contributions
of the MgI and MgII lines to the synthetic spectra at
$T_{\rm eff}$ = 8000  and 6000~K, respectively. They stress the
 importance of accurately determining line profile calculations
 to evaluate the far wings of these lines.
 The red wing of the MgI and MgII lines extends over 1000~\AA\/  until the
 CaII H/K lines at 6000~K. The contribution of the blanketing on the red and
 blue  sides is mostly due to  MgII lines, since most Mg is ionized at the
photosphere of the models shown in Figs.~\ref{MgUV8000} and ~\ref{MgUV6000}.

\begin{figure}
\resizebox{0.46\textwidth}{!}
{\includegraphics*{34067fig9.eps}}
  \caption{Comparison  of the MgHe line (red line) to the sum of the 
  two components of the Mg$^+$He  line (blue line). 
  $T = 6000$~K and  $n_{\mathrm{He}} = 1\times10^{22}$~cm$^{-3}$. \label{sig6000MgHeMgIIHe}}
\end{figure}

\begin{figure}
\centering  
\resizebox{0.46\textwidth}{!}{\includegraphics*{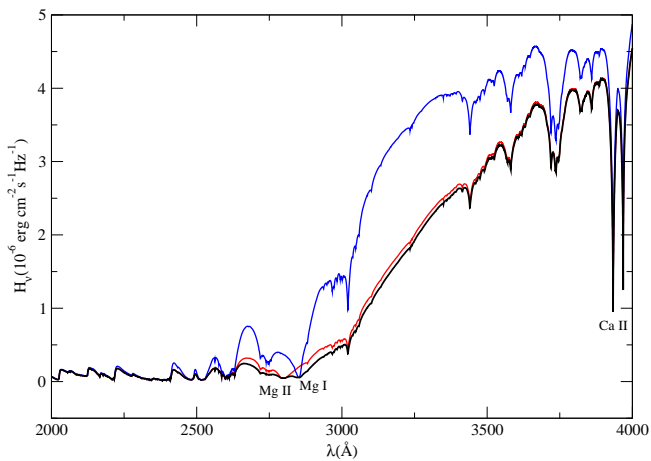}}
\caption  {Synthetic spectrum computed at $T_{\rm eff} = 6000~K$,
  $\log\,{\rm Ca/He} = -10.5$, $\log\,{\rm Fe/He} = -9.3$,
  $\log\,{\rm Mg/He} = -9.2$, and $\log\,{\rm H/He} = -3.5$,
  compared to synthetic spectra
  computed without the MgII (in blue) and the MgI (in red) line opacity.
\label{MgUV6000}}
\end{figure}

\subsection{Study of the line parameters \label{sec:parameter}}

Since in a model atmosphere calculation the resulting 
line profile is the integration of the flux in all layers from the deepest
to the uppermost, it is also important that the line centers be adequately 
represented. This means that they can be non-Lorentzian at the high densities 
of the innermost layers and Lorentzian in the upper atmosphere, but with different widths 
than predicted by the inadequate
hydrogenic van der Waals approximation that is generally used for the long-range 
interaction to calculate line cores.

The impact approximation is widely used to describe the central region of 
pressure-broadened spectral lines at low densities.
The impact theories of pressure broadening 
\citep{baranger1958a,kolb1958} are based on the assumption of sudden
collisions (impacts)  between the radiator and
perturbing atoms, and they are valid  when frequency displacements
\mbox {  $\Delta$ $\omega$ = $\omega$ - $\omega_0$} and gas densities are
sufficiently  small.
In impact broadening, the duration of the collision is assumed to be small
compared to the interval  between collisions, and the results describe the line
within a few line widths of the center.
One outcome of our unified approach is that we may evaluate
the difference between the impact limit and the general unified profile, and
establish with certainty the region of validity of an assumed Lorentzian
profile.  
The line parameters of the Lorentzian can be obtained in the impact limit 
(s $\rightarrow$ $\infty$) of the general 
calculation of the autocorrelation function (Eq.~(\ref{eq:gcl})).
 The phase shifts are then given by
Eq.~(\ref{eq:phase})  with the integral  taken between s=0 and $\infty$.
  In this way, the results described here are applicable to a more general
line profile and opacity evaluation for the same perturbers 
at any given layer in the photosphere.

\begin{figure}
\resizebox{0.46\textwidth}{!}{\includegraphics*{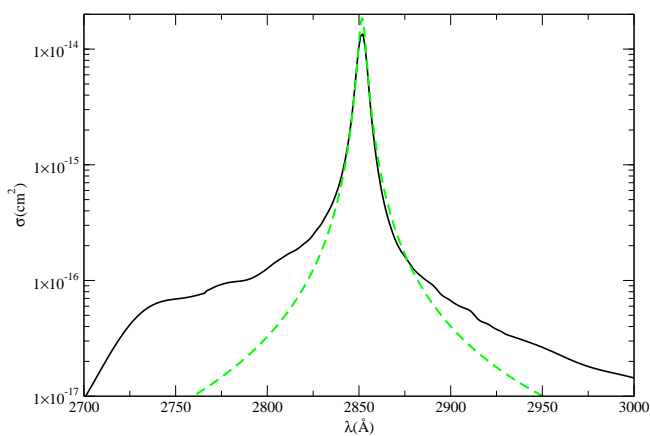}}
\caption  {Theoretical absorption cross-sections of the Mg line
  compared to the  Lorentzian profile (green dashed line)
   obtained in the impact approximation. 
The  density of the perturbers
  is $n_{\mathrm{He}} = 10^{21}$ cm$^{-3}$.
The temperature is 8000~K.
\label{rossdens1e21}}
\end{figure}

\begin{figure}
 \centering
\resizebox{0.46\textwidth}{!}
{\includegraphics*{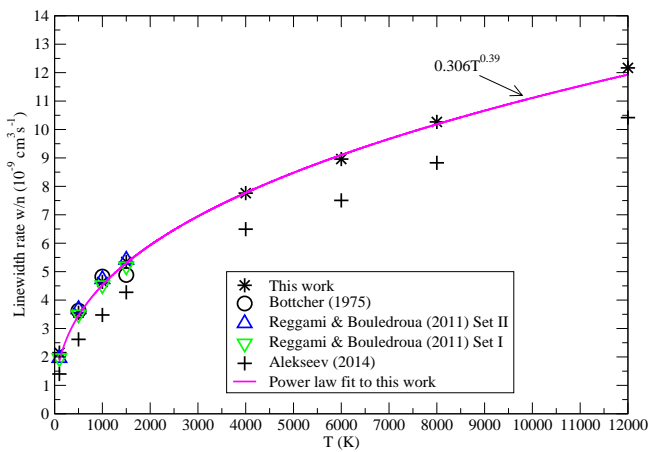}}
\caption  {Variation with temperature of the FWHM
 of the 
 resonance lines of Mg perturbed by He collisions.
 \citet{bottcher1975}(circle),  \citet{reggami2011} Set~I (green downward-pointing triangle), 
 \citet{reggami2011} Set~II (blue upward-pointing triangle), 
  Alekseev (2014, see footnote) (plus), this work (star), and a power-law fit
  (magenta line).
 The rates are in units of 10$^{-9}$ s$^{-1}$ cm$^{-3}$.}
\label{width}
\end{figure}

In the upper atmosphere of the cool white dwarfs under consideration, 
the helium atom density is on the order of $10^{21}$ cm$^{-3}$
in the region of line core formation. As the line satellites are well 
separated from the main line, we can check  that the
impact approximation is still good at this density,
 with the understanding that it will not give a correct line wing
 (Fig.~\ref{rossdens1e21}).

This  was also true for the $H$ and $K$ lines of the resonance lines of CaII 
(Fig.2 of \citet{allard2013c}).
When line satellites  are close to the parent line, as they are  for the 3s-2p line
 in He$_2$ \citep{allard2013e} or the Mg~b triplet \citep{allard2016a}, 
 the situation changes drastically 
 and leads to a complex behavior of the dependence of the line shape, 
and to the conventional line width and shift parameters on He
density. 
 
The broadening of the core of Mg lines  by helium
collisions has  never been measured in the laboratory, but other 
theoretical data have been obtained by \citet{reggami2011} 
and \citet{bottcher1975}. We report in Fig.~\ref{width} the values given 
in Table~8 of \citet{reggami2011}. They agree very well with 
the semi-classical calculations obtained using the MRCI potentials. 
To predict the impact parameters,
the intermediate and long-range part  of the potential energies 
need to be accurately determined.
The absence of a long-range part of the potentials in MOLCAS data of Alekseev (2014) 
leads to  systematically lower values of the full-width at half-maximum (FWHM). 

The dependence  of the impact line width on temperature
is presented in  Fig.~\ref{width}. The FWHM
$w$  is linearly 
dependent on the He density, and a power law in temperature given  by
\begin{equation}
w = 0.306 \times 10^{-9} n_{\mathrm{He}} \, T^{0.39}
\end{equation}
($w$ is in cm$^{-1}$,   $n_{\mathrm{He}}$ in cm$^{-3}$,  and $T$ in K)
 accurately represents the numerical results as shown in Fig.~\ref{width}.

\section{Astrophysical application to WD2216-657\label{sec:astro}}
To test our improved line profiles, we examined WD 2216-657 (LP 119-34),
a DZ star with broad Mg II resonance lines. For our analysis, we used the \textit{Gaia} DR2 parallax measurement of this
object \citep{brown2018,prusti2016}, the $VRI$
and $JHK_S$ photometry reported in \citet{subasavage2017}, the visible spectrum
obtained by \citet{subasavage2017}, and the International Ultraviolet
Explorer (IUE) spectra \citep{weidemann1989,zeidler1986}.

To fit WD 2216-657, we used the atmosphere code described in \citet{blouin2018}
and the fitting procedure outlined in \citet{dufour2007}. The solid angle,
the effective temperature, and the
Ca/He abundance ratio were fit simultaneously to the visible spectroscopy,
the UV spectroscopy, 
and the photometry using a $\chi^2$ minimization method. The abundance ratio
between the different
heavy elements was kept constant during the minimization procedure, but we
manually adjusted the abundance of C, Fe, and Mg to fit the many spectral
lines found in both the UV and the
visible spectra. The surface gravity $\log g$ was obtained from
the solid angle and the parallax measurement.

\begin{figure}
        \centering      
        \includegraphics[width=0.8\columnwidth]{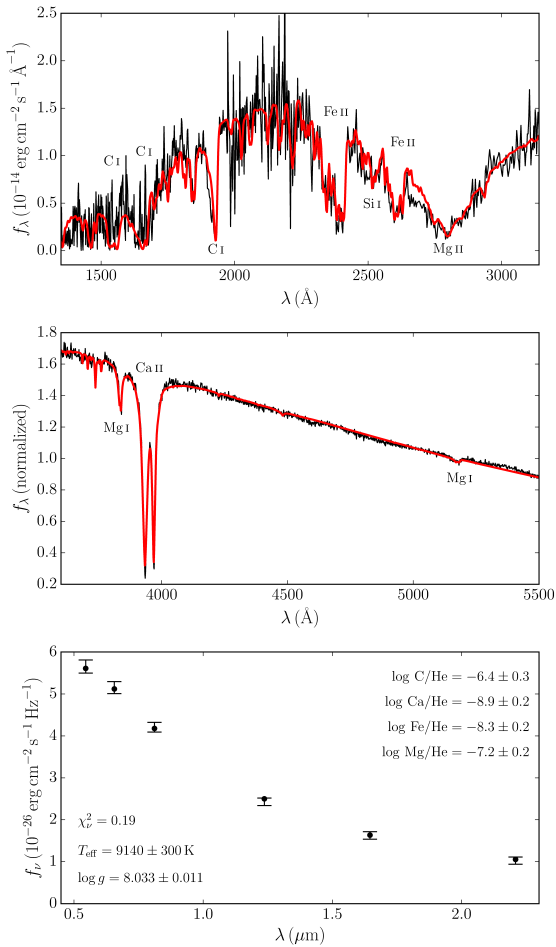}
        \caption{Our best photometric and spectroscopic fit of WD 2216-657.}
        \label{fig:WD2216-657}
\end{figure}

Figure~\ref{fig:WD2216-657} shows our best spectroscopic and photometric
fit of this object. Our solution is
in perfect agreement with the photometric observations and nicely
reproduces the C I, Fe II, Mg I, Mg II, and Si I features.
Our fitting parameters, given in the bottom panel of
Fig.~\ref{fig:WD2216-657},
are in close agreement with those previously found by \citet{wolff2002}
($T_{\rm eff}=9200 \pm 300\,{\rm K}$, $\log\,{\rm C/He}=-6.0 \pm 0.5$, $\log\,{\rm Ca/He}=-9.1$,
$\log\,{\rm Fe/He}=-7.95 \pm 0.70$ and $\log\,{\rm Mg/He}=-6.8 \pm 0.7$).
We note, however,
that \citet{wolff2002} concluded that WD 2216-657 has hydrogen in
its atmosphere, with $\log\,{\rm H/He}=-4^{+0.5}_{-1.0}$.
They reached this conclusion because they needed
to include opacity from Ly$\alpha$ broadening to explain the shape of the
blue part of the IUE spectrum.
As our hydrogen-free models are able to properly reproduce the shorter
wavelength part of the IUE spectrum, hydrogen is superfluous for our
analysis, and we therefore assume that WD 2216-657
has a hydrogen-free atmosphere. In any case, due to the non-visibility
of H$\beta$, a hydrogen abundance above 
$\log\,{\rm H/He} \approx -4.5$ is incompatible with the visible spectrum
of WD 2216-657.

Other studies \citep{subasavage2008,subasavage2017,weidemann1989} have
concluded that WD 2216-657 is hotter than what our results
suggest (they found
$T_{\rm eff}=10610 \pm 570\,{\rm K}$, $9770 \pm 660\,{\rm K,}$ and $9600 \pm 400\,{\rm K}$).
However, we found that while effective temperatures
above $T_{\rm eff} = 9400\,{\rm K}$ are compatible 
with the photometric observations, they are rejected by the
relative intensities of the Mg I and Mg II lines. 
In particular, at an effective temperature above $9400\,{\rm K}$,
the magnesium abundance required to fit the
Mg I and Mg II spectral lines implies that the Mg II 4481 line
should be visible. As this line is
not seen in the visible spectrum, we conclude that the effective
temperature of WD 2216-657 must be cooler
than $9400\,{\rm K}$.

For the purpose of this work, the most important aspect of our analysis
of WD 2216-657 is certainly our 
fit to the Mg II resonance lines. The previous detailed analysis of
these lines \citep{zeidler1986} had
two important problems. First, the Mg abundance obtained from the visible
spectrum 
($\log\,{\rm Mg/He} = -7.7$) was not compatible with the Mg abundance
required to fit the Mg II resonance lines ($\log\,{\rm Mg/He} = -6.7$).
Moreover, \citet{zeidler1986} were unable to
simultaneously fit the core and the wings of the Mg II resonance lines,
which indicates a problem with the broadening mechanism assumed for
these lines. They attempted to solve this problem by
multiplying the van der Waals broadening constant of these lines by an
arbitrary factor, but this
was still insufficient to remove the core-wings discrepancy.
Thanks to our improved line profiles,
our analysis of WD 2216-657 is not affected by any of these two problems.
The Mg/He abundance ratio that we find is compatible with both the
visible and the UV spectra, and our fit to the
Mg II resonance lines is compatible with both the core and the wings.

\section{Conclusion}
  We have performed  theoretical  calculations of the collisional profiles 
of the resonance lines of Mg  and Mg$^+$   perturbed by a hot
dense helium plasma using a unified theory of spectral line
broadening and high-quality  ab initio potentials.
While laser-produced plasmas approach the conditions of white dwarf 
atmospheres, they are transient and difficult to diagnose. 
Experimental laboratory tests are being done to determine whether traditional 
precision
experiments in an environment necessarily at lower density and temperature 
than in a white dwarf
star can verify the potentials through measurements of the wings and 
line core broadening 
in absorption spectra in the
accessible ultraviolet and visible regions.
 The  good fit of  spectra of  Ross 640 and 
 LP 658-2  acquired with the Faint Object Spectrograph on board the
 Hubble Space Telescope, which was   obtained by \citet{blouin2018},
 also confirms our opacity data.
 The spectra reported here are computed with methods that have a long
 history of validation both in astrophysical and laboratory applications.
 It is particularly significant that no parameters are adjusted adjusted
 ad hoc, and that the results are soundly based on first principles used in
 the underpinning atomic physics and atmosphere codes, enabled by expertise
 in their use.

\begin{acknowledgements}
One of us (NFA) would like to  acknowledge V.~Alekseev for  kindly 
 providing the  Mg--He molecular potentials and dipole moments 
 before publication.
  The authors thank  the referee for helpful comments
and suggestions  that  have improved the clarity of the paper.
 This work used observations made with the NASA/ESA Hubble Space
Telescope, and obtained from the Hubble Legacy Archive, which is a
collaboration between the Space Telescope Science Institute (STScI/NASA),
the Space Telescope European Coordinating Facility (STECF/ESA) and the
Canadian Astronomy Data Centre (CADC/NRC/CSA).
  Support for MAST for non-HST data is provided by the 
NASA Office of Space Science via grant NNX09AF08G and by other grants and 
contracts.
 This work has made use of data from the European Space Agency (ESA)
mission {\it Gaia} (\url{https://www.cosmos.esa.int/gaia}), processed by
the {\it Gaia} Data Processing and Analysis Consortium (DPAC,
\url{https://www.cosmos.esa.int/web/gaia/dpac/consortium}).
Funding for
the DPAC has been provided by national institutions, in particular the
institutions participating in the {\it Gaia} Multilateral Agreement.
 This work has made use of the Montreal White Dwarf
  Database \citep{mwdd2017}.

\end{acknowledgements}

\bibliographystyle{aa}
\bibliography{34067}
\end{document}